# Title:

Measurement uncertainty and unicity of single number quantities describing the spatial decay of speech level in open-plan offices.

# Authors:

Lucas LENNE[a,b], Patrick CHEVRET[a], Étienne PARIZET[b]
[a] Institut National de Recherche et de Sécurité (INRS), France
[b] Laboratory of Vibration and Acoustics (LVA), INSA Lyon, France

# Corresponding author:

Lucas LENNE: lucas.lenne@inrs.fr

# Abstract:

The ISO 3382-3 standard (2012) defines single number quantities (SNQs) which evaluate the acoustic quality of open-plan offices, but does not address the issue of measurement uncertainties. This study focusses on the SNQs present in this standard related to spatial decay of speech, i.e. $D_{2S}$, $L_{pAS4m}$ and $r_c$. The aim is to provide additional information to the limited literature on the measurement uncertainties of these SNQs by use of both analytical developments and a stochastic approach based on simulations. The accuracy of the analytical developments was studied thanks to simulations of the sound propagation within a series of offices (1 layout, 16 acoustic configurations with different screen heights and different acoustic qualities of screens and ceiling). The SNQs obtained in the simulations cover a wide range: $D_{2S}$ between 3.4 and 7.5 dB(A), $L_{pAS4m}$ between 40.6 and 51.9 dB(A) and $r_c$ between 2.5 and 14.7 m. Therefore, the simulations are representative of a broad set of acoustic qualities. Estimated uncertainties have a magnitude of 0.4 dB(A) for $D_{2S}$ and vary between 0.4 and 0.7 dB(A) for $L_{pAS4m}$ and between 0.2 and 1.5m for $r_c$ over a measurement line comprising 7 measurement positions. The simulations also raise the question of describing the acoustic quality of an office using a single value for the indicators. The results of the simulations show that in some cases, $D_{2S}$ values significantly depend on the measurement line, leading to a strong increase of its measurement uncertainty if a unique value is to be considered.

# Keywords:

# 1 Introduction

## 1.1 Context of the study

Noise constitutes a major source of annoyance in open-plan offices. Among all noise sources in this type of work environment, conversational noise is the most detrimental, especially when speech is intelligible [1]. Therefore, the question of the propagation of speech noise is a major aspect of the acoustic quality of open-plan offices. The ability of an office (including layout and furnishings) to limit this propagation of speech is estimated according to the ISO 3382-3 measurement standard which is currently undergoing revision at the time of writing this paper [2]. This estimation is made thanks to single number quantities (SNQs) which consist in the aggregation of octave band values from 125 Hz to 8 kHz. The ISO 3382-3 standard defines SNQs reflecting two different philosophies with regard to the acoustic quality of open-plan offices. The first one is based on the description of the spatial decay of the A-weighted sound pressure level (SPL) of a signal presenting a speech spectrum, while the second approach focuses on the spatial decay of speech intelligibility.

This study focusses on the first approach because its SNQs are widely used in national standards [3,4].

## 1.2 Single number quantities

The approach of ISO 3382-3, which focusses on the spatial decay of the A-weighted SPL of a signal presenting a speech-like spectrum (later referred to as speech), defines two SNQs:

- The decrease of the A-weighted SPL of speech by doubling the distance to the source: $D_{2S}$,
- The A-weighted SPL of speech at the distance of 4 m from the source: $L_{pAS4m}$.

At the time of this study, the ISO 3382-3 standard was being revised in view to introducing a third SNQ inspired by the work of Nilsson and Hellström [5]. This SNQ, called the comfort distance and noted $r_c$, is defined as the distance from the source where the A-weigthed SPL of speech falls below 45 dB(A).

These three SNQs are derived from the linear regression of the A-weighted SPL of speech as a function of the base-2 logarithm of the distance to the source; the $D_{2S}$ being the slope of the regression, $L_{pAS4m}$ the intercept at 4 meters, and $r_c$ calculated from these two quantities (Figure 1).

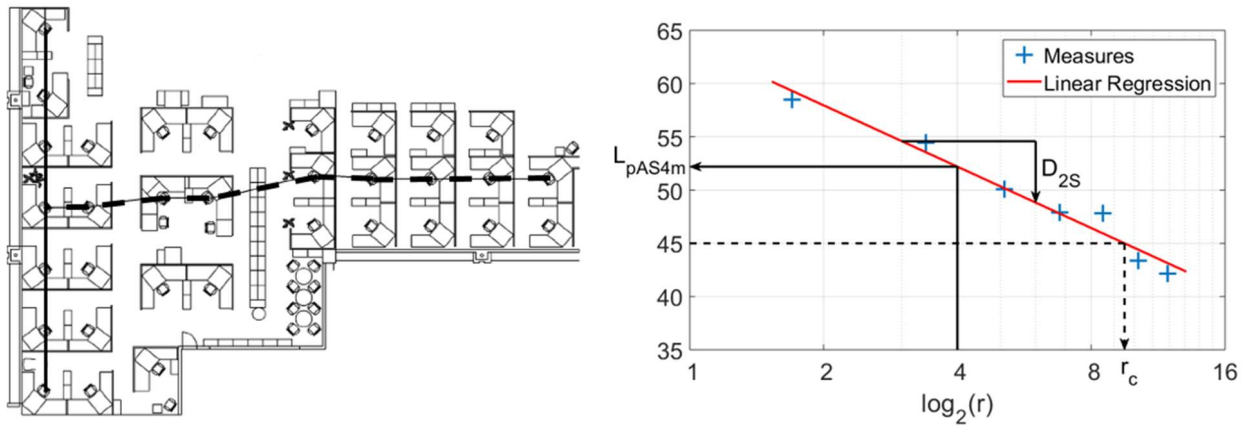

*Figure 1: Examples of measurement paths (solid and dashed lines) from ISO 3382-3 (2012) (left) and the evaluation of the SNQs (right).*

In the field, the measurement procedure consists in measuring the SPL along a virtual line from one workstation to another. The measurement path must be as straight as possible, but small deviations are allowed if the configuration of the workplace does not allow drawing a perfect line. During the measurement, an omnidirectional loudspeaker is placed at one end of the path and the A-weigthed sound pressure levels ($L_{pASi}$) and distances from the sources ($r_i$) are measured at each workstation along the path. The SNQs are then calculated using equation 1 (for a path composed of N workstations).

$$D_{2S} = -\frac{N \cdot \sum L_{pASi} \log_2(r_i) - \sum L_{pASi} \sum \log_2(r_i)}{N \cdot \sum \log_2(r_i)^2 - (\sum \log_2(r_i))}$$

$$L_{pAS4m} = \frac{1}{N}\sum L_{pASi} + D_{2S} \cdot \frac{1}{N}\sum \log_2\left(\frac{r_i}{4}\right) \quad (1)$$

$$r_c = 4 \cdot 2^{\frac{L_{pAS4m}-45}{D_{2S}}}$$

## 1.3 Aim of the study

At the time of writing the paper, the current version of the ISO 3382-3 standard was undergoing revision and does not mention measurement uncertainties, an essential aspect of the evaluation. The lack of consideration of measurement uncertainties makes it impossible to assess the relevance of target values set by national and international standards. Indeed, if the measurement uncertainty is large compared to the difference between two office qualities, the target values lose their discriminatory aspect.

Very few studies in the literature address the issue of the measurement uncertainties of the SNQs of ISO 3382-3. First, Haapakangas *et al.* [6], who studied the relation between various quantities (including the SNQs of the ISO 3382-3 standard) and the perceived disturbance by noise, reported that the measurement uncertainties of $D_{2S}$ and $L_{pAS4m}$ are 1 dB(A) and 1.5 dB(A), respectively. These values are based on "the experience and unpublished data of the authors". Then, Yadav and colleagues [7] conducted an experimental study of the repeatability of the metrics of ISO 3382-3. Based on 36 measurements, the study concluded that the measurement uncertainty of $D_{2S}$ and $L_{pAS4m}$ are 0.6 dB(A) and 1.0 dB(A), respectively. Finally, Hongisto *et al.* [8] conducted a round-robin test, evaluating the reproducibility of the metrics of ISO 3382-3 (2012). The test took part in an office of rather low acoustic quality ($D_{2S}$, $L_{pAS4m}$ and $r_c$ of about 3.8 dB(A), 52.5 dB(A) and 17 m). The test resulted in a measurement uncertainty for a single measurement path comprised between 0.2 and 0.5 dB(A) for $D_{2S}$, between 0.9 and 1.3 for $L_{pAS4m}$ and between 1.9 and 2.4 m for $r_c$.

The aim of this study is to provide useful additional results based on analytical developments and Monte-Carlo simulations. These results could be used to improve the current version of the ISO 3382-3 standard.

Analytical expressions can be obtained from the law of propagation of uncertainty as described by the International Bureau of Weights and Measures [9]. They are based on linearizations so their accuracy has to be assessed. To that end, a set of simulations mimicking measurements made in various acoustic configurations in an open-plan office were performed.

The question of reporting a unique value for each SNQ in a single acoustic area – a single acoustic area is a space where both the ceiling and furniture design are homogeneous - should also be considered in the standard. In its current version, at least two measurement paths in each acoustic area are prescribed and the two SNQs have to be reported, but the question of a mean value is still being debated. Therefore, the second aim of the current study is to assess the necessity of performing several measurements within one acoustic area. On the one hand, if the SNQs are constant within an area, carrying out a single measurement of the SNQs and evaluating their measurement uncertainties would be enough to characterise its acoustic quality. On the other hand, if the SNQs vary within the area, it would be necessary to perform several measurements of the SNQs. This question is addressed by performing numerical simulations.

This paper first presents the analytical developments leading to expressions of the measurement uncertainties of the SNQs. Secondly, a numerical approach is presented and applied to a case of an open-plan office which includes several acoustic configurations. Thirdly, an evaluation of the accuracy of the analytical expressions is conducted, based on the same case study. Finally, the results of the simulations are used to investigate the unicity of the SNQs within an acoustic area.

## 2 Analytical expressions of the uncertanties

The International Bureau of Weights and Measures issued guidelines for the evaluation of measurement uncertainties [9]. In particular, this guide describes the propagation of uncertainties, which states that if the measurand is evaluated indirectly (using secondary quantities and an equation linking these quantities and the measurand), the measurement uncertainties associated with intermediate quantities contribute to the uncertainty in the evaluation of the measurand.

In the case of a measurand $Y$ evaluated using N quantities $\vec{X} = (X_i)_{1 \leq i \leq N}$ of uncertainty $(u_{X_i})_{1 \leq i \leq N}$ and whose errors are independent, the measurement uncertainty of the measurand can be evaluated using equation 2.

$$u_Y^2 = \sum_{i=1}^{N} \left( \frac{\partial Y}{\partial X_i} \bigg|_{\vec{X}} \cdot u_{X_i} \right)^2 \qquad (2)$$

However, this expression is the result of a linearization and should therefore be used with caution if the relation between the measurand and any intermediary quantity is strongly non-linear or if the measurement uncertainties are large.

## 2.1 Source of uncertainty in the measurement of SNQs

The three SNQs considered in this paper are evaluated from the linear regression of the level as a function of the logarithm in base 2 of the distance from the source at workstations on a path across the office. Therefore, the measurement uncertainty of the SNQs ($u_{SNQ}$) can be evaluated using equation 3 below, where $u_{L_{pASi}}$ and $u_{r_i}$ correspond respectively to the uncertainty of the A-weighted SPL and to that of the distance to the source of the i$^{th}$ measurement position on the path.

$$u_{SNQ}^2 = \sum_{i=1}^{N} \left( \frac{\partial SNQ}{\partial L_{pASi}} \cdot u_{L_{pASi}} \right)^2 + \sum_{i=1}^{N} \left( \frac{\partial SNQ}{\partial r_i} \cdot u_{r_i} \right)^2 \qquad (3)$$

With respect to the measurement uncertainty of sound pressure levels, the ISO 3382-3 standard states that the microphones used for the measurement must fulfil the same specifications as Class 1-sound level meters, as defined by IEC 61672-1 (2003) [10]. These specifications partly relate to measurement uncertainties. It is therefore possible to use this standard to determine the measurement uncertainties of levels in the seven octave bands from 125 Hz to 8000 Hz (see Table 1).

| $f_c$ (Hz)    | 125 | 250 | 500 | 1000 | 2000 | 4000 | 8000 |
|---------------|-----|-----|-----|------|------|------|------|
| $u_{oct}$ (dB) | 0.9 | 0.9 | 0.8 | 0.8  | 0.9  | 1.2  | 1.8  |

*Table 1: Measurement uncertainty of octave band levels deduced from the characterization of Class 1 sound level meters in the IEC 61672-1 (2003) standard.*

From these measurement uncertainties of the seven octave-band levels, the measurement uncertainty of the A-weighted SPL ($u_{L_{pAS}}$) is evaluated according to the propagation of uncertainties, which results in equation 4, where $L_{pAS}$ is the A-weighted SPL, $L_{p,oct}$ is the octave band levels and $A_{oct}$ corresponds to the A-weighting.

$$u_{L_{pAS}}^2 = \sum_{oct=1}^{7} \left( \frac{\partial L_{pAS}}{\partial L_{p,oct}} \cdot u_{oct} \right)^2 = \frac{1}{\left( 10^{L_{pAS}/10} \right)^2} \sum_{oct=1}^{7} \left( 10^{\frac{L_{p,oct} + A_{oct}}{10}} \cdot u_{oct} \right)^2 \qquad (4)$$

Concerning the distances from the source, the measurement uncertainty of the instrumentation is assumed to be in the range of a few tenths of a millimetre, whatever the measuring device (laser rangefinder or measuring tape). From the practical viewpoint, the measurement conditions are not ideal in an open-plan office: most often, there are objects (panels, PC screens, desk lamps, etc.)

between the microphone and the loudspeaker. It may therefore be difficult to place one end of the measuring tape at the microphone and the other at the loudspeaker, while keeping it straight (or aiming at the microphone with the laser rangefinder from the loudspeaker). Therefore, the distances are measured between two points located above the microphone and the loudspeaker, which greatly increases the measurement uncertainty. Instead of a few tenths of a millimetre, the measurement uncertainty of distances ($u_r$) can be increased up to a few centimetres. For the rest of the study, it will be considered to be 5 cm, meaning that the measurement error is less than 10 cm in 95% of cases.

Besides the errors due to the measurement of octave-band levels and distances, another source of uncertainty is related to the positioning of the measurement apparatus. Indeed, ISO 3382-3 states that the measurements must be carried out by positioning the sound source and the microphone at the position of the head of a person at the workstation. This defines the positioning of the measurement equipment fairly well. However, two persons carrying out the measurement will not place the instrumentation exactly at the same location. It is therefore important to consider this positioning error, which induces an error in the evaluation of:

- The distance between the source and the workstation at which the microphone is theoretically positioned.
- The level at the workstation at which the microphone is theoretically positioned.

In the present study, it is assumed that the positioning error follows a normal distribution in both horizontal directions and that the apparatus is placed within a square with 20-cm sides, centred at its theoretical position 95% of the time. On the one hand, these two assumptions result in an uncertainty for the distance from the source to the microphone ($u_{r,pos}$) of 6.3 cm that must be added to the distance measurement uncertainty. On the other hand, the relation between the positioning error and the induced error in SPL is complex due to the presence of the layout and the acoustic treatments of the office. This relation would however be measurable *in situ,* but it would make more sense to carry out a direct empirical evaluation of measurement uncertainties. Furthermore, the acoustic pressure field is assumed homogenous in the close proximity of the workstations. Therefore, the error in the evaluation of the level caused by the positioning error is neglected.

## 2.2  Analytical expressions of the measurement uncertainty

The analytical expressions for the measurement uncertainties of the SNQs are obtained by applying linearization equations 3 and 4 to the expression of the SNQs (Eq. 1). The results of these analytical developments (presented in the annexe of this paper) are the following expressions:

$$u_{D_{2S}}^2 = \frac{\text{Cov}\left(\log_2(r), \left(\log_2(r) - \overline{\log_2(r)}\right) \cdot u_{L_{pAS}}^2\right) + \frac{u_r^2}{\log(2)^2} \cdot \text{Cov}\left(\alpha, \overline{\alpha - \bar{\alpha}}/r^2\right)}{N \cdot \text{Var}(\log_2(r))^2} \qquad (5)$$

$$u_{L_{pAS4m}}^2 = \frac{\overline{u_{L_{pAS}}^2}}{N} + \frac{D_{2S}^2 \cdot u_r^2}{N \cdot \log(2)^2} \cdot \overline{\left(\frac{1}{r^2}\right)} + \overline{\log_2\left(\frac{r}{4}\right)^2} \cdot u_{D_{2S}}^2 - \frac{2}{N} \cdot \overline{\log_2\left(\frac{r}{4}\right)} \cdot \frac{\text{Cov}\left(\log_2(r), u_{L_{pAS}}^2\right)}{\text{Var}(\log_2(r))}$$
$$- \frac{2 \cdot D_{2S} \cdot u_r^2}{N \cdot \log(2)^2} \cdot \overline{\log_2\left(\frac{r}{4}\right)} \cdot \frac{\text{Cov}\left(\alpha, 1/r^2\right)}{\text{Var}(\log_2(r))} \qquad (6)$$

$$u_{r_c}^2 = \left(\frac{\log(2) \cdot r_c}{D_{2S}}\right)^2 \cdot \left[u_{L_{pAS4m}}^2 + \left(\log_2\left(\frac{r_c}{4}\right)^2 - 2 \cdot \log_2\left(\frac{r_c}{4}\right) \cdot \overline{\log_2\left(\frac{r}{4}\right)}\right) \cdot u_{D_{2S}}^2\right] \qquad (7)$$

$$+\frac{2 \cdot Cov\left(\log_2(r), u^2_{L_{pAS}}\right) + \frac{D_{2S} \cdot u^2_r}{\log(2)^2} \cdot Cov\left(\alpha, {1}/{r^2}\right)}{N \cdot Var(\log_2(r))} \cdot \log_2\left(\frac{r_C}{4}\right)\Bigg]$$

where the quantity α is defined as:

$$\alpha_i = L_{pASi} + 2 \cdot D_{2S} \cdot \log_2(r_i)$$

In these expressions, the measurement uncertainties of the SNQs are proportional to the inverse of the square root of the number of measurement positions, meaning that increasing the number of workstations on the measurement path will reduce the uncertainty.

It is noteworthy that the evaluation of these expressions from field measurements remains quite complex. A theoretical analysis of the orders of magnitude of the different terms of these expressions seems difficult at this stage. However, on the basis of the measurements of the SNQs carried out by INRS in 13 offices (36 paths in total, unpublished results), it appears that:
 - the first term of $D_{2S}$ gives a correct uncertainty value to within 0.06 dB(A);
 - the first two terms of $L_{pAS4m}$ provide a good approximation of the uncertainty to within 0.06 dB(A);
 - and the first two terms of $r_C$ provide a good approximation to within 0.3 m.

More measurements should be made in order to consolidate these findings.

Despite this difficulty of evaluating these expressions, reporting the measurement uncertainties should be a recommendation of the ISO 3382-3 standard. One suggestion could be to provide a calculation tool based on the previous expressions to the users of this standard.

## 3   Numerical simulations

The analytical expressions (5-7) are based on approximations whose accuracy has to be evaluated. To that end, a numerical approach was used with the help of the RayPlus v8.1.0 software. This software is based on a ray tracing method. It takes into account reflections by surfaces, the transmission loss of partitions and diffractions by edges [11,12]. Results are given in octave bands between 125 Hz and 8000 Hz. For the simulations presented hereafter, the receivers (microphones) are spherical cells 5 cm in diameter and, in order to obtain the convergence of the calculated levels to within less than 0.1 dB, 10 million rays are traced for each source.

### 3.1   Case study

The office designed for the simulations has a simple geometrical layout and is intended to be representative of call centres. In this type of office, people have a predominantly individual activity, mainly on the telephone. They do not need to communicate with each other. This is why the layout is homogeneous with partitions installed between the workstations facing each other. The office is 13.6 m by 8.5 m and accommodates 32 workstations. The workstations are laid out in four rows separated by a 2 m wide centre aisle (see Figure 2). The two sidewalls present windows along their entire length. In the office, four measurement paths were defined, as shown on Figure 2. The paths are numbered from P1 to P4. The office constitutes a single acoustic area, in accordance with the recommendations of ISO 3382-3, since both the acoustic treatment and the layout are similar throughout the office.

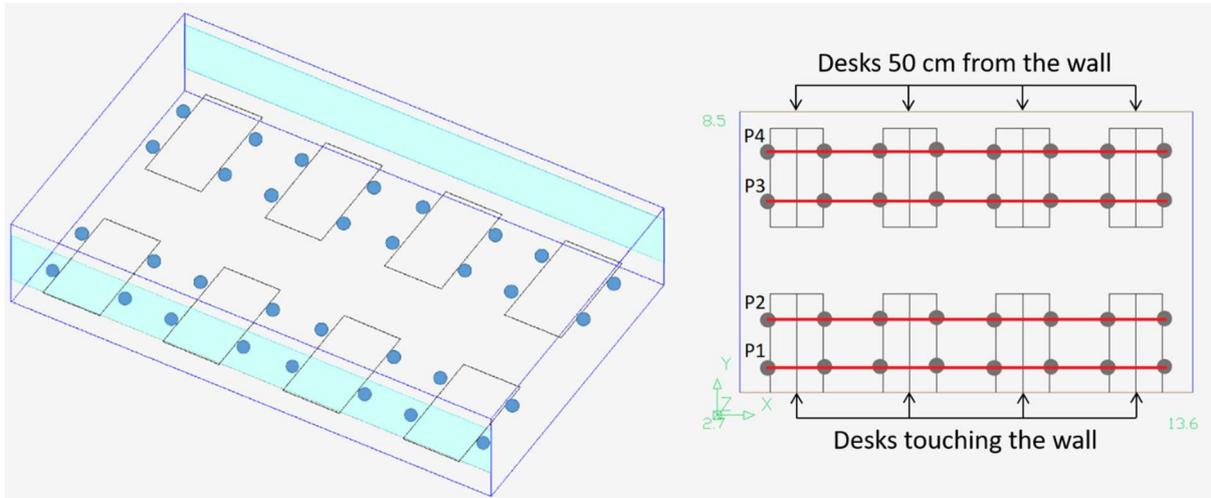

*Figure 2: Layout of the simulated office with 32 workstations (grey dots). Four measurement paths (red lines) numbered from P1 (bottom) to P4 (top) are defined in accordance with the ISO3382-3 standard.*

In the simulations, 16 furniture configurations were defined including different acoustic properties of the ceiling, different heights and acoustic properties of the screens. For the ceiling, two classes of material were considered: class A ($\alpha_w \geq 0.9$) and class C ($0.6 \leq \alpha_w < 0.9$) according to the ISO 11654 (1997) standard [13]. The screen height was set to four different values: 190 cm (H1), 150 cm (H2), 130 cm (H3) and 110 cm (H4). For each height, the same classes of acoustic material were considered as for the ceiling. The suspended ceiling is described by an absorption coefficient for each octave band, while the acoustic screens are described by an absorption coefficient and a transmission loss (TL) value for each octave band. The acoustic properties corresponding to the lowest requirement of both the class A ceiling (C1) and screens (S1) and the class C ceiling (C2) and screens (S2) are presented in Figure 3.

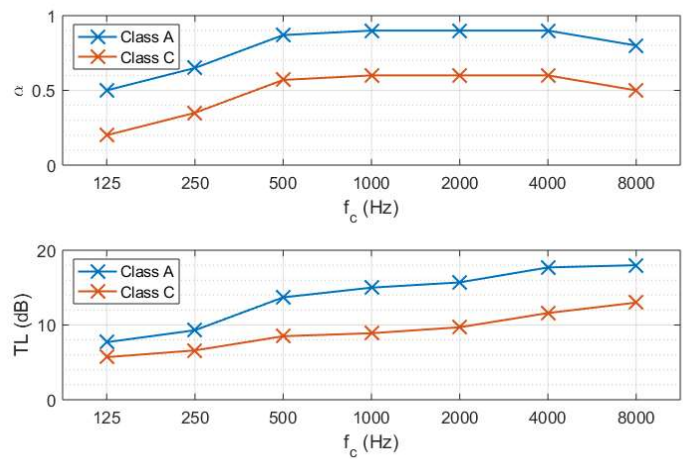

*Figure 3: Acoustic characteristics used for the simulations. Top : absorption properties of the ceiling and the screens. Bottom : transmission losses of screens.*

In the following, the 16 acoustic configurations of the office are labelled using the combination of the numbers corresponding to the height of the screens, the acoustic quality of the ceiling and that of the screens. For example, configuration "312" corresponds to the office in which the screens are 130 cm high (H3), the ceiling is of class A (C1) and the screens of class C (S2).

## 3.2 Uncertainties calculation procedure

The simulations enable the evaluation of the measurement uncertainties by applying the Monte-Carlo method, as described by the International Bureau of Weights and Measures [14]. This method consists in emulating a series of measurements by randomly selecting errors. As said previously, the sources of error come from the positioning and the measuring instruments. These two types are treated differently.

To enable the emulation of the positioning errors, nine positions of the apparatus were defined at each workstation within the simulation. The nine positions were set on a 3-by-3 20-cm wide square grid centred on the theoretical position of the workstation. The simulation results for each measurement position and for each octave band on 81 levels, correspond to possible pairs of source and microphone positions.

The first step of the process is to emulate the positioning of the apparatus by the person performing the measurement. The positioning error of each measurement device is selected randomly (1.a in Figure 4) and follows a bivariate normal distribution, as described previously. The levels estimated for the selected positions of the source and microphones are interpolated from the results of the simulation (1.b in Figure 4). At this stage, the process does not take into account the measurement uncertainty of the apparatus.

In the second step, measurement errors are selected randomly (2.a in Figure 4). On the one hand, the measurement error of octave-band SPL follows a uniform law in the interval $\left[-u_{oct}\sqrt{3}, u_{oct}\sqrt{3}\right]$, which is common practice when the measurement uncertainty is related to tolerance limits. On the other hand, the distance measurement error follows a centred normal distribution whose variance $u_r^2$ is equal to $0.05^2$. These measurement errors are then added to the outcomes of the first step of the process, resulting in the emulation of a complete measurement of the SNQs (2.b, in Figure 4).

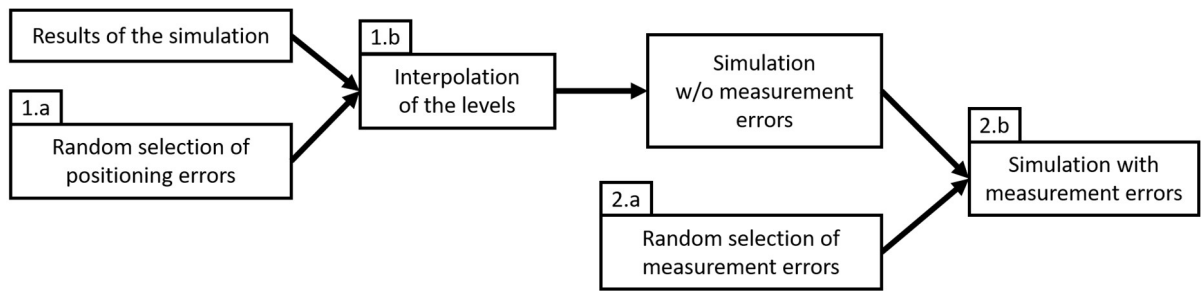

*Figure 4: Process of the Monte-Carlo approach.*

The whole process is then repeated several times in order to provide the measurement uncertainties. The number of repetitions was gradually increased until convergence, i.e. until the magnitude of the measurement uncertainties obtained by the different runs were within 0.01 dB for $D_{2S}$ and $L_{pAS4m}$ and within 1 cm for $r_c$. This criterion was fulfilled with 10000 repetitions.

The mean values of the SNQs and their uncertainties (i.e. standard deviation) were calculated from the 10000 runs for each path and each acoustic configuration after verifying the normality of the distributions.

# 4 Results

The raw results of the simulations are presented in Figure 5, which shows the distribution of SNQ values for the 4 paths in the 16 acoustic configurations (each histogram summarizes $16 \times 4 \times 10^4 = 6.4 \cdot 10^5$ values). $D_{2S}$ varies from 3.4 to 7.5 dB(A), $L_{pAS4m}$ from 40.6 to 51.9 dB(A) and $r_c$ from 2.5 to 14.7 m. They are representative of a wide variety of open-plan offices. With respect to the classification of open-plan offices of the German VDI 2569 standard [4], these simulated values range from unclassified (bad acoustic quality) to class B (good office for administrative work, suitable for a call centre). They also cover the complete range of values recommended by the French NF S31-199 (2016) and the international standard ISO/FDIS 22955 (2020) [15] dedicated to the acoustic quality of open-plan offices.

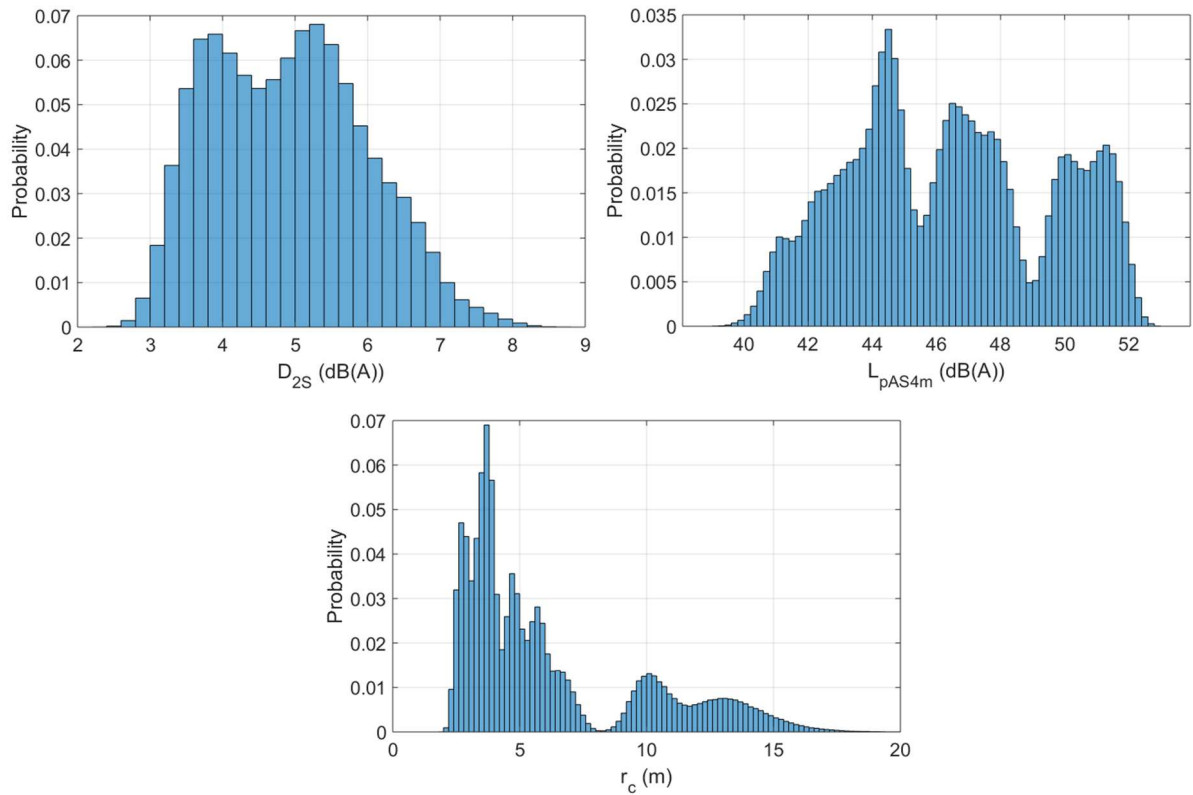

*Figure 5: Distribution of the SNQs for the complete set of acoustic configurations and measurements paths.*

## 4.1 Measurement uncertainties of the case study

The measurement uncertainties are represented in Figure 6, and their overall extremums are presented in Table 2 together with the measurement uncertainties put forward by Haapakangas *et al.* [6], Yadav *et al.* [7] and Hongisto *et al.* [8].

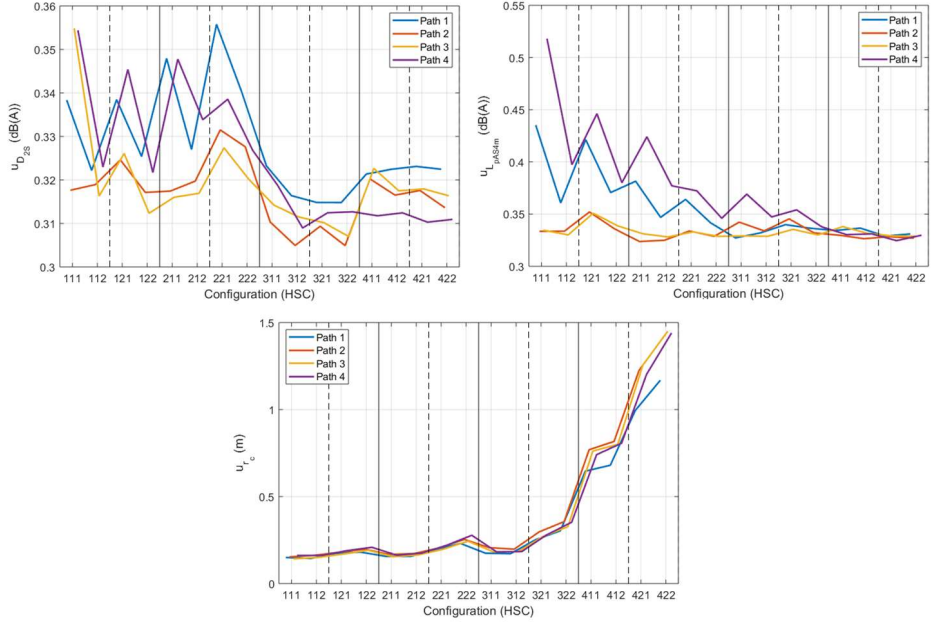

*Figure 6: Magnitudes of the measurement uncertainties of the SNQs estimated from the simulations, for each office configuration and each measurement path.*

|  | Simulations |  | Haapakangas (2017) | Yadav (2018) | Hongisto (2021) |
| --- | --- | --- | --- | --- | --- |
|  | N = 7 | N = 5 | N = ? | N = 5-8 | N = 5 |
| $u_{D2S}$ (dB(A)) | 0.4 | 0.4-0.5 | 1 | 0.61 | 0.2-0.5 |
| $u_{LpAS4m}$ (dB(A)) | 0.4-0.6 | 0.4-0.7 | 1.5 | 1.04 | 0.9-1.3 |
| $u_{rc}$ (m) | 0.2-1.5 | 0.2-2.2 | - | - | 1.9-2.4 |

*Table 2: Magnitude of measurement uncertainties obtained with the simulations (considering the measurement paths in their entirety – N=7 – or only the 4 closest measurement positions – N=4) compared to the literature.*

While the simulated measurement uncertainties are of the same order of magnitude as the values presented in the literature, their direct comparison is made difficult because the magnitude of the measurement uncertainty of the SNQs greatly depends on the number of measurement points along the measurement path. To illustrate this, the simulations were also performed on each path by considering only the five measurement positions closest to the source. These results are also presented in Table 2.

In agreement with equations 5 to 7, the uncertainty increased as the number of measurement points decreased. Table 2 shows that even for the shortest measurement paths (N=5), the simulated uncertainties are smaller than those in the literature. The magnitudes of the simulated measurement uncertainty are very close to those of Hongisto *et al.*

### 4.2 Accuracy of the analytical expressions

The assessment of the accuracy of the analytical expressions derived in section 2.2 can be performed for the current case by comparing the values obtained analytically to those obtained with the Monte-Carlo approach. The comparisons are presented in figure 7 for each path and for the 16 configurations. In this figure, the blue crosses represent the exact estimation of the uncertainties and the red circles, the values rounded up to the next one-tenth, which is the usual practice for measurement uncertainties.

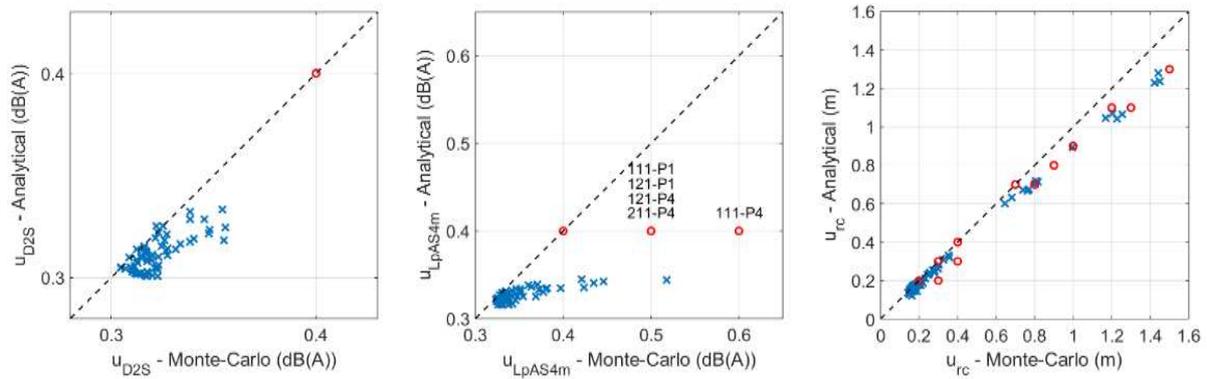

*Figure 7: Comparison of the measurement uncertainty estimated by the analytical and numerical approach for each measurement path and each acoustic configuration - exact estimated values (crosses) and rounded up to the nearest one-tenth (circles).*

In this figure, it appears that the measurement uncertainty of $D_{2S}$ is estimated accurately by the analytical expression since both approaches give the same value of 0.4 dB for any measurement path of all the acoustic configurations. The same can be said for the measurement uncertainty of $r_c$, for which the difference between the two approaches remains below 0.2 m, reaching this value only when the measurement uncertainty is large (over 1.2 m).

Concerning the estimation of the measurement uncertainty of $L_{pAS4m}$, the two approaches give the same results for 59 of the 64 measurement paths. Differences appear in three configurations (paths 1 and 4 of configurations 111, 121 and path 4 of configuration 211). In these three cases where the dividers are high, the analytical expressions underestimate the measurement uncertainty by up to 0.2 dB(A).

This underestimation, which remains quite low, could be explained by the hypothesis that the error in the positioning of the measurement apparatus does not induce significant variations in the SPL measured.

This is illustrated in Figure 8 which represents the differences between the sound pressure fields for two positions of the source of measurement path 4 in the acoustic configuration 111. The map, derived from simulations with Rayplus, shows that a small difference in the source position (separation distance of 28 cm between the two sources) can change the propagation in the office: the presence of propagation paths for S2 (prevented by the first acoustic screen for S1) results in variations of several decibels at workstations 3, 4 and 7. These variations could explain the increase in the measurement uncertainty of the SNQs that are not taken into account by the analytical expressions.

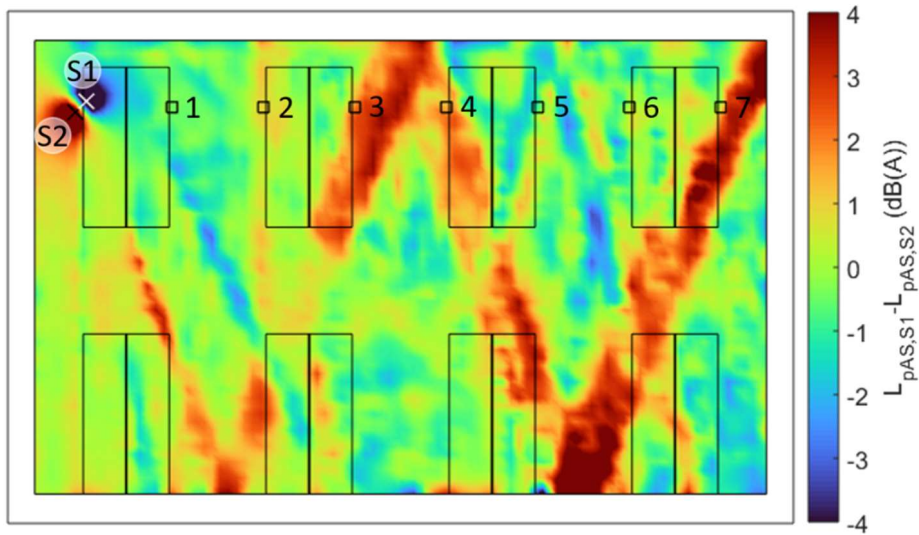

*Figure 8: Difference of the sound pressure field for two positions of the source (path 4, configuration 111). The crosses represent the position of the source. Measurement positions are represented by squares.*

### 4.3 Unicity of SNQs

The question of the possibility of a unique SNQs in a given acoustic area is of prime importance because it enables defining in a standard the recommended method to use for reporting data. The 2012 version of ISO 3382-3 requires considering at least two measurement paths in an acoustic area. The VDI 2569 standard is more complete and requires that the measurement be carried out on a number of paths determined by the number of workstations in the acoustic area, but this question is still under debate [7], especially in the framework of the revision of ISO 3382-3. In this respect, simulations are of precious help because they enable emulating the measurement of SNQs and uncertainties using several measurement paths in offices with a wide range of acoustic qualities.

In our case, the effect of the measurement path on SNQs is illustrated by Figure 9, which represents the mean values and the 95% confidence interval of the SNQs for each measurement path in each acoustic configuration. It seems that the choice of the path may have a significant influence on $D_{2S}$ only: for example, using paths 2 and 4 give different measurements of $D_{2S}$ (the 95% confidence intervals do not overlap).

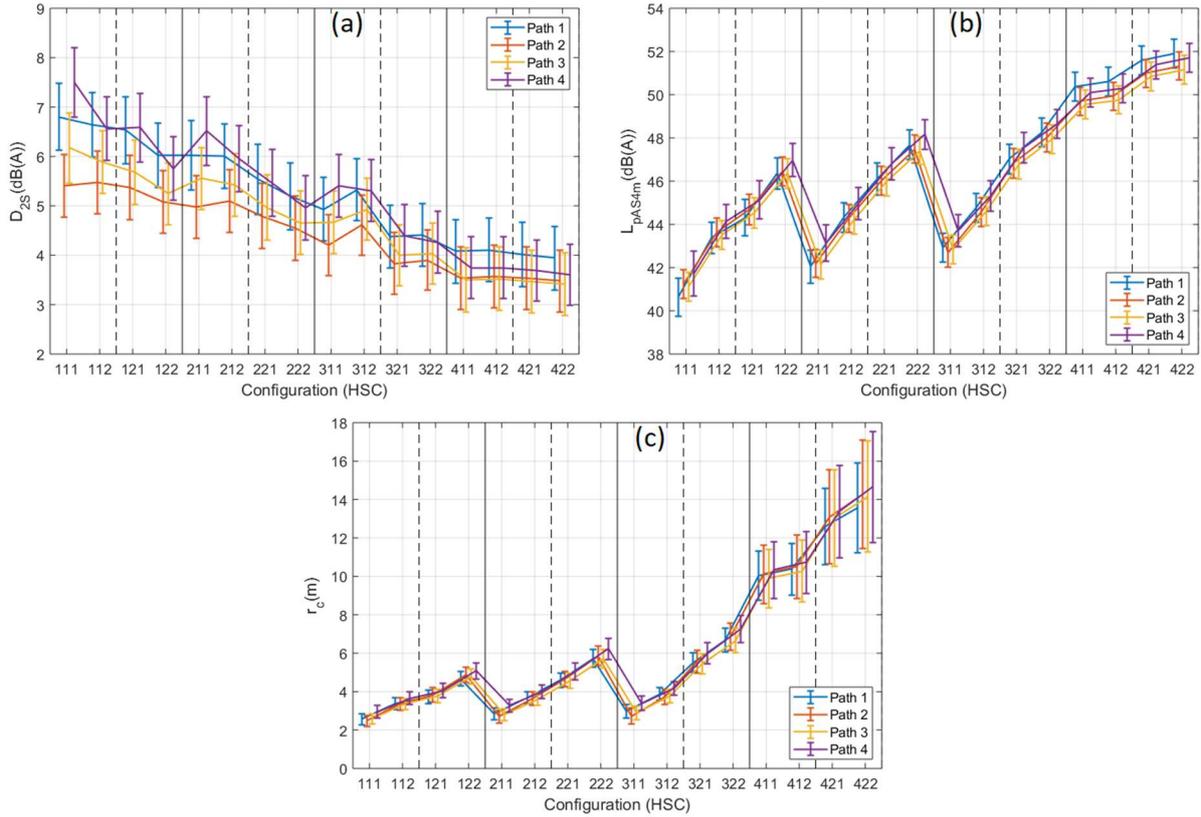

*Figure 9: Mean values and 95% confidence interval of $D_{2S}$ (a), $L_{pAS4m}$ (b) and $r_c$ (c) for the four measurement paths in each of the 16 acoustic configurations of the simulated office.*

However, if one wishes to characterize an office with a unique value of the SNQs, one should evaluate the uncertainty from the entire office (and not from a given measurement path). This was calculated for all the configurations of the case study and gave values between 0.4 and 0.9 dB(A) for $D_{2S}$, 0.5 and 0.6 dB(A) for $L_{pAS4m}$ and 0.3 and 1.5 m for $r_c$. For example, the measurement uncertainty of $D_{2S}$ in configuration 111 becomes 0.9 dB(A) instead of 0.4 dB(A) for a single path. Therefore, the 95% confidence interval of $D_{2S}$ becomes ±1.8 dB(A), which is not accurate enough for the evaluation of the acoustic quality of the office.

Therefore, the use of a unique value of $D_{2S}$ seems to be questionable, whereas there is no reason to raise doubts concerning the unicity of $L_{pAS4m}$ and $r_c$ in the office studied. These observations on the unicity of the SNQs in the office are in line with those made by Yadav and colleagues [7] who found, for adjacent measurement paths (called Type 2 error in the paper), that $L_{pAS4m}$ does not differ significantly between paths, whereas $D_{2S}$ may take significantly different values. However, firm conclusions could not be drawn due to the small number (n=7) of Type 2 measurements conducted in the study.

The explanation for the non-unicity of a $D_{2S}$ value in a single acoustic area is the same as that given previously for the difference between the analytical expression of the uncertainties and the Monte-Carlo approach: the spatial decay of speech depends on the measurement paths because changing the position of the source can have a significant effect on the propagation map and greatly change the measured values from one path to another. By way of example, Figure 10 shows two maps of SPL in the office when the source is located at the beginning of paths 1 and 2, respectively, in configuration 111 (190-cm-high class A screens and class A ceiling). In this configuration, $D_{2S}$ is evaluated at 6.8±0.7 dB(A) on path 1 and 5.4±0.6 dB(A) on path 2. The source of path 1 is more enclosed than that of path 2, resulting in few contributions from the source to the distant measurement positions. In the case of the second measurement path, the sound pressure field at measurement positions 3, 4, 5 and 7 is dominated by reflections from the windows on both sides of the office.

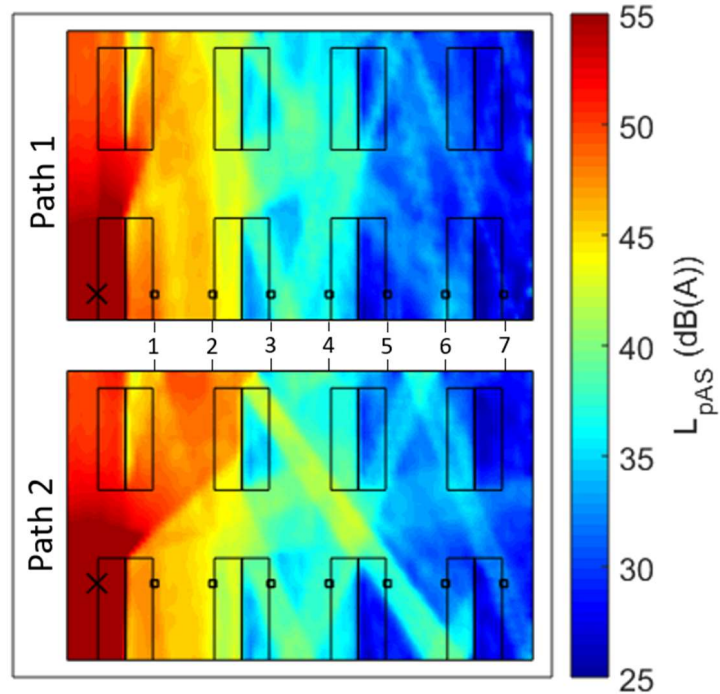

Figure 10: SPL maps illustrating the path dependency of $D_{2S}$.

# 5  Conclusion

The aim of this study was to address the uncertainties of acoustic SNQs used to evaluate the quality of an open-plan office. This issue is a major aspect which is not considered in the current version of ISO 3382-3 at the time of conducting this study.  In this paper, analytical developments based on the recommendation of the International Bureau of Weights and Measures were carried out. The resulting expressions of the measurement uncertainties are still quite complex, but they could be used easily in addition to the existing standard or introduced in a revision of its current version. Since they are based on linearization, these expressions were checked for accuracy as the SNQs present non-linearities with respect to the measurement intermediates (distances to the source and SPLs).

Therefore, a set of simulations were carried out. These simulations emulated a single open-plan office with a wide range of acoustic configurations resulting in a wide range of values for the SNQs considered. From the simulations, it appeared that the expressions of measurement uncertainties established in this paper were very precise and more accurate for $D_{2S}$ and $r_c$ than for $L_{pAS4m}$. For the latter, the analytical expression gave the same results as the simulations for 59 of the 64 measurements considered, and underestimated the measurement uncertainty by less than 0.2 dB(A) for the other five.

The question of the unicity of SNQs within an acoustic area was also addressed on the base of simulations. The results indicated that it seems necessary to report several measurements of the SNQs in a single acoustic area of an open-plan office (as required by the ISO 3382-3 (2012) standard). Indeed, even within these areas, where the acoustic fittings were homogeneous, the SNQs, and more particularly $D_{2S}$, can vary significantly from one measurement path to another.

## Annex: Analytical development of measurement uncertainties

In the equations, the sums, means, variances and covariances are evaluated over the measurement positions.

### Measurement uncertainty of $D_{2S}$

$$D_{2S} = -\frac{N \cdot \sum L_{pAS} \cdot \log_2(r) - \sum L_{pAS} \cdot \sum \log_2(r)}{N \cdot \sum \log_2(r)^2 - (\sum \log_2(r))^2}$$

$$u_{D_{2S}}^2 = \sum \left(\frac{\partial D_{2S}}{\partial L_{pASi}} \cdot u_{L_{pASi}}\right)^2 + \sum \left(\frac{\partial D_{2S}}{\partial r_i} \cdot u_r\right)^2 = u_{D_{2S}}^2(L) + u_{D_{2S}}^2(r)$$

### Evaluation of $u_{D_{2S}}^2(L)$

$$\frac{\partial D_{2S}}{\partial L_{pASi}} = -\frac{N \cdot \log_2(r_i) - \sum \log_2(r)}{N \cdot \sum \log_2(r)^2 - (\sum \log_2(r))^2} = -\frac{\log_2(r_i) - \overline{\log_2(r)}}{N \cdot \text{Var}(\log_2(r))}$$

$$u_{D_{2S}}^2(L) = \frac{\sum (\log_2(r_i) - \overline{\log_2(r)})^2 u_{L_{pASi}}^2}{N^2 \cdot \text{Var}(\log_2(r))^2} = \frac{\text{Cov}\left(\log_2(r), (\log_2(r) - \overline{\log_2(r)})\right) \cdot u_{L_{pAS}}^2}{N \cdot \text{Var}(\log_2(r))^2}$$

### Evaluation of $u_{D_{2S}}^2(r)$

$$\frac{\partial D_{2S}}{\partial r_i} = \frac{\partial \log_2(r_i)}{\partial r_i} \cdot \frac{\partial D_{2S}}{\partial \log_2(r_i)} = \frac{-1}{\log(2) \cdot r_i} \cdot (T_1 - T_2)$$

$$T_1 = \frac{(N \cdot L_{pASi} - \sum L_{pAS})(N \cdot \sum \log_2(r)^2 - (\sum \log_2(r))^2)}{(N \cdot \sum \log_2(r)^2 - (\sum \log_2(r))^2)^2} = \frac{(N \cdot L_{pASi} - \sum L_{pAS})}{N^2 \cdot \text{Var}(\log_2(r))}$$

$$T_2 = \frac{2(N \cdot \log_2(r_i) - \sum \log_2(r))(N \cdot \sum L_{pAS} \log_2(r) - \sum L_{pAS} \cdot \sum \log_2(r))}{(N \cdot \sum \log_2(r)^2 - (\sum \log_2(r))^2)^2}$$

$$T_2 = \frac{-2 \cdot D_{2S} \cdot (N \cdot \log_2(r_i) - \sum \log_2(r))}{N^2 \cdot \text{Var}(\log_2(r))}$$

$$\frac{\partial D_{2S}}{\partial r_i} = \frac{-1}{\log(2) \cdot r_i} \cdot \frac{(N \cdot L_{pASi} - \sum L_{pAS}) + 2 \cdot D_{2S} \cdot (N \cdot \log_2(r_i) - \sum \log_2(r))}{N^2 \cdot \text{Var}(\log_2(r))}$$

$$\frac{\partial D_{2S}}{\partial r_i} = \frac{-1}{\log(2) \cdot r_i} \cdot \frac{N \cdot \alpha_i - \sum \alpha}{N^2 \cdot \text{Var}(\log_2(r))} = \frac{-1}{\log(2) \cdot r_i} \cdot \frac{\alpha_i - \overline{\alpha}}{N \cdot \text{Var}(\log_2(r))}$$

$$u_{D_{2S}}^2(r) = \frac{u_r^2}{\log(2)^2 \cdot N^2 \cdot \text{Var}(\log_2(r))^2} \cdot \sum \frac{(\alpha_i - \overline{\alpha})^2}{r_i^2} = \frac{u_r^2 \cdot \text{Cov}\left(\alpha, {\alpha - \overline{\alpha}}/{r^2}\right)}{\log(2)^2 \cdot N \cdot \text{Var}(\log_2(r))^2}$$

$$u_{D_{2S}}^2 = \frac{\text{Cov}\left(\log_2(r), (\log_2(r) - \overline{\log_2(r)})\right) \cdot u_{L_{pAS}}^2 + \frac{u_r^2}{\log(2)^2} \cdot \text{Cov}\left(\alpha, {\alpha - \overline{\alpha}}/{r^2}\right)}{N \cdot \text{Var}(\log_2(r))^2}$$

## Measurement uncertainty of L<sub>pAS4m</sub>

$$L_{pAS4m} = \frac{1}{N} \sum L_{pAS} + D_{2S} \cdot \frac{1}{N} \cdot \sum \log_2\left(\frac{r}{4}\right)$$

$$u_{L_{pAS4m}}^2 = \sum \left(\frac{\partial L_{pAS4m}}{\partial L_{pASi}} \cdot u_{L_{pASi}}\right)^2 + \sum \left(\frac{\partial L_{pAS4m}}{\partial r_i} \cdot u_r\right)^2 = u_{L_{pAS4m}}^2(L) + u_{L_{pAS4m}}^2(r)$$

### Evaluation of $u_{L_{pAS4m}}^2(L)$

$$\frac{\partial L_{pAS4m}}{\partial L_{pASi}} = \frac{1}{N} + \frac{\partial D_{2S}}{\partial L_{pASi}} \cdot \overline{\log_2\left(\frac{r}{4}\right)}$$

$$u_{L_{pAS4m}}^2(L) = \frac{1}{N^2} \sum u_{L_{pAS}}^2 + \overline{\log_2\left(\frac{r}{4}\right)}^2 \cdot \sum \left(\frac{\partial D_{2S}}{\partial L_{pASi}} \cdot u_{L_{pASi}}\right)^2 + \frac{2}{N} \cdot \overline{\log_2\left(\frac{r}{4}\right)} \cdot T_3$$

$$T_3 = \sum \frac{\partial D_{2S}}{\partial L_{pASi}} \cdot u_{L_{pASi}}^2 = \frac{-1}{N \cdot \text{Var}(\log_2(r))} \sum (\log_2(r_i) - \overline{\log_2(r)}) \cdot u_{L_{pASi}}^2$$

$$T_3 = -\frac{\text{Cov}\left(\log_2(r), u_{L_{pAS}}^2\right)}{\text{Var}(\log_2(r))}$$

$$u^2_{L_{pAS4m}}(L) = \overline{\frac{u^2_{L_{pAS}}}{N}} + \overline{\log_2\left(\frac{r}{4}\right)^2} \cdot u^2_{D_{2S}}(L) - \frac{2}{N} \cdot \overline{\log_2\left(\frac{r}{4}\right)} \cdot \frac{Cov\left(\log_2(r), u^2_{L_{pAS}}\right)}{Var(\log_2(r))}$$

## Evaluation of $u^2_{L_{pAS4m}}(r)$

$$\frac{\partial L_{pAS4m}}{\partial r_i} = \frac{D_{2S}}{N} \cdot \frac{1}{\log(2) \cdot r_i} + \frac{\partial D_{2S}}{\partial r_i} \cdot \overline{\log_2\left(\frac{r}{4}\right)}$$

$$u^2_{L_{pAS4m}}(r) = \left(\frac{D_{2S} \cdot u_r}{N \cdot \log(2)}\right)^2 \cdot \sum \frac{1}{r_i^2} + \overline{\log_2\left(\frac{r}{4}\right)}^2 \cdot \sum \left(\frac{\partial D_{2S}}{\partial r_i} \cdot u_r\right)^2 + \frac{2 \cdot D_{2S} \cdot u_r^2}{N \cdot \log(2)} \cdot \overline{\log_2\left(\frac{r}{4}\right)} \cdot T$$

$$T_4 = \sum \frac{\partial D_{2S}}{\partial r_i} \cdot \frac{1}{r_i} = \frac{-1}{\log(2) \cdot N \cdot Var(\log_2(r))} \cdot \sum \frac{\alpha_i - \overline{\alpha}}{r_i^2}$$

$$T_4 = \frac{-1}{\log(2)} \cdot \frac{Cov(\alpha, 1/r^2)}{Var(\log_2(r))}$$

$$u^2_{L_{pAS4m}}(r) = \frac{D^2_{2S} \cdot u_r^2}{N \cdot \log(2)^2} \cdot \overline{\left(\frac{1}{r^2}\right)} + \overline{\log_2\left(\frac{r}{4}\right)}^2 \cdot u^2_{D_{2S}}(r) - \frac{2 \cdot D_{2S} \cdot u_r^2}{N \cdot \log(2)^2} \cdot \overline{\log_2\left(\frac{r}{4}\right)} \cdot \frac{Cov(\alpha, 1/r^2)}{Var(\log_2(r))}$$

$$u^2_{L_{pAS4m}} = \overline{\frac{u^2_{L_{pAS}}}{N}} + \frac{D^2_{2S} \cdot u_r^2}{N \cdot \log(2)^2} \cdot \overline{\left(\frac{1}{r^2}\right)} + \overline{\log_2\left(\frac{r}{4}\right)}^2 \cdot u^2_{D_{2S}} - \frac{2}{N} \cdot \overline{\log_2\left(\frac{r}{4}\right)} \cdot \frac{Cov\left(\log_2(r), u^2_{L_{pAS}}\right)}{Var(\log_2(r))}$$
$$- \frac{2 \cdot D_{2S} \cdot u_r^2}{N \cdot \log(2)^2} \cdot \overline{\log_2\left(\frac{r}{4}\right)} \cdot \frac{Cov(\alpha, 1/r^2)}{Var(\log_2(r))}$$

## Measurement uncertainty of $r_c$

$$r_c = 4 \cdot 2^{\frac{L_{pAS4m}-45}{D_{2S}}} = 4 \cdot \exp\left(\log(2) \cdot \frac{L_{pAS4m} - 45}{D_{2S}}\right)$$

$$u^2_{r_c} = \sum \left(\frac{\partial r_c}{\partial L_{pASi}} \cdot u_{L_{pASi}}\right)^2 + \sum \left(\frac{\partial r_c}{\partial r_i} \cdot u_r\right)^2 = u^2_{r_c}(L) + u^2_{r_c}(r)$$

## Evaluation of $u_{r_c}^2(L)$

$$\frac{\partial r_c}{\partial L_{pASi}} = \frac{\log(2) \cdot r_c}{D_{2S}} \cdot \frac{\partial L_{pAS4m}}{\partial L_{pASi}} - \frac{\log(2) \cdot r_c}{D_{2S}} \cdot \log_2\left(\frac{r_c}{4}\right) \cdot \frac{\partial D_{2S}}{\partial L_{pASi}}$$

$$\frac{\partial r_c}{\partial L_{pASi}} = \frac{\log(2) \cdot r_c}{D_{2S}} \cdot \left(\frac{\partial L_{pAS4m}}{\partial L_{pASi}} - \log_2\left(\frac{r_c}{4}\right) \cdot \frac{\partial D_{2S}}{\partial L_{pASi}}\right)$$

$$u_{r_c}^2(L) = \left(\frac{\log(2) \cdot r_c}{D_{2S}}\right)^2 \cdot \left[u_{L_{pAS4m}}^2(L) + \log_2\left(\frac{r_c}{4}\right)^2 \cdot u_{D_{2S}}^2(L) - 2 \cdot \log_2\left(\frac{r_c}{4}\right) \cdot T_5\right]$$

$$T_5 = \sum \frac{\partial L_{pAS4}}{\partial L_{pASi}} \cdot \frac{\partial D_{2S}}{\partial L_{pASi}} \cdot u_{L_{pASi}}^2 = \sum \left(\frac{1}{N} + \frac{\partial D_{2S}}{\partial L_{pASi}} \cdot \overline{\log_2\left(\frac{r}{4}\right)}\right) \cdot \frac{\partial D_{2S}}{\partial L_{pASi}} \cdot u_{L_{pASi}}^2$$

$$T_5 = \frac{T_3}{N} + \overline{\log_2\left(\frac{r}{4}\right)} \cdot u_{D_{2S}}^2(L)$$

$$u_{r_c}^2(L) = \left(\frac{\log(2) \cdot r_c}{D_{2S}}\right)^2 \cdot \left[u_{L_{pAS4m}}^2(L) + \left(\log_2\left(\frac{r_c}{4}\right)^2 - 2 \cdot \log_2\left(\frac{r_c}{4}\right) \cdot \overline{\log_2\left(\frac{r}{4}\right)}\right) \cdot u_{D_{2S}}^2(L)\right]$$
$$+ \left(\frac{\log(2) \cdot r_c}{D_{2S}}\right)^2 \cdot \frac{2}{N} \cdot \log_2\left(\frac{r_c}{4}\right) \cdot \frac{Cov\left(\log_2(r), u_{L_{pAS}}^2\right)}{Var(\log_2(r))}$$

## Evaluation of $u_{r_c}^2(r)$

$$\frac{\partial r_c}{\partial r_i} = \frac{\log(2) \cdot r_c}{D_{2S}} \cdot \left(\frac{\partial L_{pAS4m}}{\partial r_i} - \log_2\left(\frac{r_c}{4}\right) \cdot \frac{\partial D_{2S}}{\partial r_i}\right)$$

$$u_{r_c}^2(r) = \left(\frac{\log(2) \cdot r_c}{D_{2S}}\right)^2 \cdot \left[u_{L_{pAS4m}}^2(r) + \log_2\left(\frac{r_c}{4}\right)^2 \cdot u_{D_{2S}}^2(r) - 2 \cdot \log_2\left(\frac{r_c}{4}\right) \cdot T_6\right]$$

$$T_6 = \sum \frac{\partial L_{pAS4}}{\partial r_i} \cdot \frac{\partial D_{2S}}{\partial r_i} \cdot u_r^2 = \sum \left(\frac{D_{2S}}{N \cdot \log(2) \cdot r_i} + \frac{\partial D_{2S}}{\partial r_i} \cdot \overline{\log_2\left(\frac{r}{4}\right)}\right) \cdot \frac{\partial D_{2S}}{\partial r_i} \cdot u_r^2$$

$$T_6 = \frac{D_{2S} \cdot u_r^2 \cdot T_4}{N \cdot \log(2)} + \overline{\log_2\left(\frac{r}{4}\right)} \cdot u_{D_{2S}}^2(r)$$

$$u_{r_c}^2(r) = \left(\frac{\log(2) \cdot r_c}{D_{2S}}\right)^2 \cdot \left[u_{L_{pAS4m}}^2(r) + \left(\log_2\left(\frac{r_c}{4}\right)^2 - 2 \cdot \log_2\left(\frac{r_c}{4}\right) \cdot \overline{\log_2\left(\frac{r}{4}\right)}\right) \cdot u_{D_{2S}}^2(r)\right]$$

$$+ \left(\frac{\log(2) \cdot r_c}{D_{2S}}\right)^2 \cdot \frac{D_{2S} \cdot u_r^2}{N \cdot \log(2)^2} \cdot \log_2\left(\frac{r_c}{4}\right) \cdot \frac{Cov\left(\alpha, 1/r^2\right)}{Var(\log_2(r))}$$

$$u_{r_c}^2 = \left(\frac{\log(2) \cdot r_c}{D_{2S}}\right)^2 \cdot \left[u_{L_{pAS4m}}^2 + \left(\log_2\left(\frac{r_c}{4}\right)^2 - 2 \cdot \log_2\left(\frac{r_c}{4}\right) \cdot \overline{\log_2\left(\frac{r}{4}\right)}\right) \cdot u_{D_{2S}}^2\right.$$

$$\left. + \frac{2 \cdot Cov\left(\log_2(r), u_{L_{pAS}}^2\right) + \frac{D_{2S} \cdot u_r^2}{\log(2)^2} \cdot Cov\left(\alpha, 1/r^2\right)}{N \cdot Var(\log_2(r))} \cdot \log_2\left(\frac{r_c}{4}\right)\right]$$